\documentclass[%
 reprint,
 amsmath,amssymb,
 aps,
 prl,
]{revtex4-1}

\usepackage[T1]{fontenc}
\usepackage{hyperref}
\usepackage{graphicx,bm,xcolor,cleveref}

\begin{document}


\title{Quantum anti-quenching of radiation from laser-driven structured plasma channels}
\author{F. Mackenroth$^{1,2}$}
\email[corresponding author: ]{mafelix@pks.mpg.de}
\author{Z. Gong$^{3,4}$}
\author{A. V. Arefiev$^{2}$}
\affiliation{$^1$Max Planck Institute for the Physics of Complex Systems, 01187 Dresden, Germany}
\affiliation{$^2$University of California at San Diego, La Jolla, CA 92093, USA}
\affiliation{$^3$KLNPT, KLHEDP and CAPT, School of Physics, Peking University, Beijing 100871, China}
\affiliation{$^4$Center for High Energy Density Science, The University of Texas at Austin, Austin, TX 78712, USA}

\date{\today}

\begin{abstract}

We demonstrate that in the interaction of a high-power laser pulse with a structured solid-density plasma-channel, clear quantum signatures of stochastic radiation emission manifest, disclosing a novel avenue to studying the quantized nature of photon emission. In contrast to earlier findings we observe that the total radiated energy for very short interaction times, achieved by studying thin plasma channel targets, is significantly larger in a quantum radiation model as compared to a calculation including classical radiation reaction, i.e., we observe quantum anti-quenching. By means of a detailed analytical analysis and a refined test particle model, corroborated by a full kinetic plasma simulation, we demonstrate that this counter-intuitive behavior is due to the constant supply of energy to the setup through the driving laser. We comment on an experimental realization of the proposed setup, feasible at upcoming high-intensity laser facilities, since the required thin targets can be manufactured and the driving laser pulses provided with existing technology.

\end{abstract}

\maketitle

Accelerated charges emit electromagnetic radiation which naturally reduces the emitting particle' energy \cite{LandauII,Jackson_1999}. Recent technological and theoretical developments of high-power laser facilities \cite{Mourou_etal_2006,Danson_etal_2015} brought acceleration regimes into grasp in which the radiative energy loss of a laser-driven electron (mass $m$, charge $e<0$) becomes strong enough to instantaneously affect its dynamics, an effect labeled instantaneous radiation reaction (RR) \cite{DiPiazza_etal_2012}. Studies of RR are receiving ever growing attention \cite{Mackenroth_etal_2013,Vranic_etal_2014,Wang_etal_2015,Arran_etal_2019} with pioneering experimental tests realized \cite{Cole_etal_2018,Poder_etal_2018} and further campaigns planned at current \cite{VulcanLaser,Hooker_etal_2008,Leemans_etal_2010,ELI_WhiteBook,Zou_etal_2015,Gales_etal_2018,CORELS} or planned \cite{Kawanaka_etal_2016,XCELS} laser facilities. 
Classically, RR is modelled as a continuous drag on an accelerated particle, accounting for the radiative energy loss, and was found to significantly alter the dynamics of various laser-driven particles \cite{Harvey_Marklund_2012,Heinzl_etal_2015,Chen_etal_2010,Tamburini_etal_2010}.
Quantum RR, on the other hand, is interpreted as a stochastic sequence of incoherent photon emissions, each depleting the electron's energy \cite{DiPiazza_etal_2010,DiPiazza_etal_2020}, reducing to the classical result for small emitted photon energies \cite{Mackenroth_DiPiazza_2013}.
\begin{figure}[b]
\centering
\includegraphics[width=\linewidth]{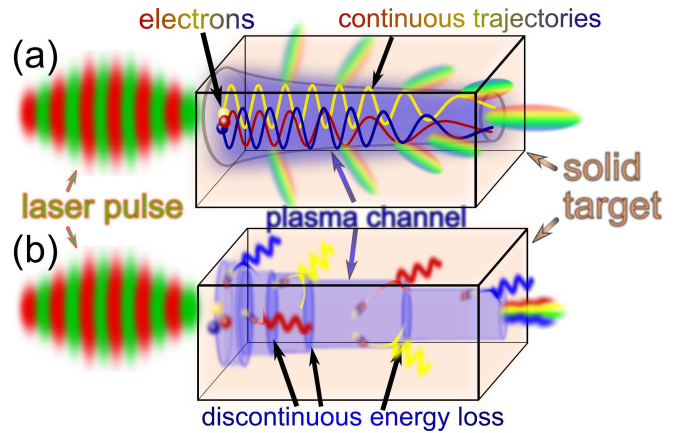}
\caption{Schematic of the studied setup of electrons driven to emit radiation (rainbow color) by an ultra-intense laser pulse (red-green) through a solid plasma channel target (orange). The emission cone (blue) decreases in size continuously according to a classical emission model (a) and stochastically according to a discrete quantum emission model (b).}
\label{fig:schematic}
\end{figure}
Coherent emissions become significant only for high particle energies 
\cite{Dinu_etal_2016}. Nevertheless, stochastic quantum effects can significantly alter RR signatures as compared to a classical model, notable in distortions of the emitting electrons' \cite{Neitz_DiPiazza_2013} or emitted photons' \cite{Blackburn_etal_2014} spectra. 

Earlier studies of quantum RR in collisions of highly relativistic electrons with ultra-intense laser pulses \cite{DiPiazza_2008,Sokolov_etal_2010,Thomas_etal_2012,Baird_etal_2019} analyzed spectral \cite{Koga_etal_2005,Li_etal_2014,Vranic_etal_2016}, as well as angular signatures of RR \cite{DiPiazza_etal_2009}. Consequently, elaborately optimized experimental tests were proposed \cite{Hammond_2010,Blackburn_2015} and large-scale plasma simulations performed, including an approximate account for stochastic QED effects \cite{Ji_etal_2014,Vranic_etal_2014,Wang_etal_2015,Duff_etal_2018}. However, as only higher-order moments of the electrons' distribution function exhibit signatures of quantum RR \cite{Niel_etal_2018}, effects of stochasticity are challenging to discriminate. One comparatively easily discriminable signature of stochastic radiation, was found in the lower quantum emission for short interaction times, since an electron emits discrete photons only after a finite interaction time \cite{Harvey_etal_2017}.
While the required sub-cycle, ultra-intense laser pulses are thus far unavailable, laser-driven plasma channels are also prolific sources of high-energy photons \cite{Stark_etal_2016,Gong_etal_2018}, are significantly affected by RR \cite{Wistisen_etal_2018,Wistisen_etal_2019,Gong_etal_2019}, and can be manufactured at low thickness \cite{GeneralAtomics}, yielding the required ultra-short interaction times.
Considering that the emission of high-energy photons clearly differs between classical and quantum regimes \cite{Harvey_etal_2009,Mackenroth_DiPiazza_2011} such plasma-channel targets should be ideal test grounds for quantum RR experiments. 

In this work we investigate the radiation yield from a laser-driven plasma channel and show how the characteristics of this emission differ in a classical and quantum electrodynamics framework. Thus we provide a clear path towards an experimental characterization of stochastic signatures in quantum RR measurements. The suggested setup is a simple solid target of varying length, structured into an inner lower-density channel surrounded by a higher-density bulk (s. Fig. \ref{fig:schematic}). Hitting this structured target with a high-intensity laser pulse turns the inner channel into a relativistically underdense plasma inside which the laser can propagate almost unperturbed \cite{Stark_etal_2016}. During its propagation the laser accelerates electrons which, in turn, emit high-energy radiation over an interaction time which is finely tunable through varying the channel length. According to the predictions of quantum quenching, one should naively expect there to be less emission predicted in a quantum analysis as compared to a classical one. Detecting the emitted radiation as function of channel length and opening angle of the emission cone this conjecture can be tested experimentally.
\begin{figure}[t]
\centering
\includegraphics[width=\linewidth]{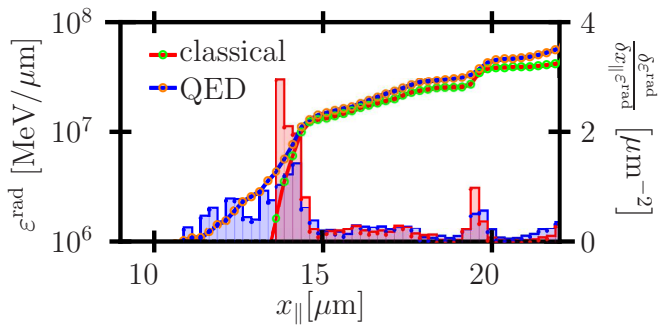}
\caption{(lines) Total emitted energy $\varepsilon^\text{rad}$ per unit transverse length (unresolved in the simulation) for varying plasma channel lengths $x_\|$ for a classical (red-green) and QED (blue-orange) emission model obtained from 2D PIC simulations (left scale). (bars) Relative increase in emitted energy per unit channel length (right scale).}
\label{fig:figure2}
\end{figure}

To test this conjecture here we ran two-dimensional particle-in-cell (PIC) simulations of a laser linearly polarized along $x_\text{las}$, propagating along $x_\|$ with wavelength $\lambda=1\,\mu$m (frequency $\omega=2\pi c/\lambda$) and intensity $I_L \approx 6\times 10^{23}$ W/cm$^2$ (peak field $E_L\approx 2.3\times 10^{13}$ V/cm) driving the relativistically underdense inner plasma channel with density $n=2 n_\text{cr}$, where $n_\text{cr}$ is the critical plasma density \cite{Piel_2014}, in a simulation box of size $70\times24\,\mu$m and a grid size of $25$ nm. We numerically collected the total energy $\varepsilon^\text{rad}$ above a detection threshold $\varepsilon^\text{det} = 100$ MeV radiated per transverse unit length, not resolved in our simulation, into a cone of angles $\theta = 15^\circ$ half-opening angle around the laser's propagation axis for differing channel lengths. We find that for large channel lengths, i.e., long interaction times the emitted radiation is approximately equal, as the quenching effect vanishes, but for short interaction times, the quantum framework predicts significantly more emission than the classical simulation (s. Fig.~\ref{fig:figure2}). Furthermore, we find the classical emission to exhibit a pronounced jump in the relative increase of the emitted energy, quantified as the energy difference $\delta \varepsilon^\text{rad}=\varepsilon^\text{rad}_{i+1}-\varepsilon^\text{rad}_i$ between two adjacent data points $i$ and $i+1$, divided by the longitudinal length step $\delta x_\| = x_\|^{i+1} - x_\|^i$ and the average energy $\varepsilon^\text{rad} = (\varepsilon^\text{rad}_{i+1}+\varepsilon^\text{rad}_i)/2$. The longitudinal location of the jump distinguishes a channel length only beyond which the classical model predicts emission into the chosen observation cone and which can be used as distinct detection signal. This appears to be in stark contrast to the previously observed suppression of radiation emission in the quantum framework due to quenching \cite{Harvey_etal_2018}. In order to explain this seeming contradiction a detailed understanding of the electrons' dynamics and their emission is required. Notably, we are going to find that the key difference to the quenching of quantum emission in the collision of an electron with a laser is the continuing driving of the radiating electrons by the co-propagating laser pulse inside the plasma channel in the present case. 
%

In previous studies it was demonstrated by means of 3D and 2D particle-in-cell (PIC) simulations and analytical models that in laser driven plasma channels strong 
currents $j_0$ 
drive quasi-static magnetic fields \cite{Gong_etal_2018,Gong_etal_2019}
, that due to ion migration electric charge separation fields are quickly compensated \cite{Jansen_etal_2018,Wang_etal_2019} and that the plasma's relativistic transparency leads to a stable laser propagation at approximately the speed of light over long distances \cite{Gong_etal_2018,Gong_etal_2019}. Hence, the ensuing electron dynamics can be well modelled by a test particle model in the combined laser and quasi-static magnetic plasma fields. In the following analysis we employ units with $\hbar =c=1$ unless stated otherwise. 
We model the channel's current density to be uniform and hence its magnetic field to be cylindrically symmetric around the channel axis and linearly increasing with distance $r$ from it. Its electrodynamic potential can then be written as ${\bm{a}}_\text{mag} = \left|m\alpha/e\right| (r/\lambda)^2 {\bm{e}}_\|$, where the dimensionless amplitude $\alpha = e\pi \lambda^2 j_0/m$ characterizes the channel magnetic field. Inside this channel magnetic field, the motion of a relativistic electron with initial energy (longitudinal momentum) $\varepsilon_0$ ($p_{\|,0}$) is confined to transverse displacements smaller than the \textit{magnetic boundary} \cite{Gong_etal_2019} $x_\text{las} \leq x_{MB} := \lambda \sqrt{(\varepsilon_0 - p_{\|,0})/(m \alpha)}$ \cite{Gong_etal_2019}. Since this condition confines the electron's transverse motion to smaller excursions than a laser's typical focal width and the laser's phase fronts inside the channel are near-flat \cite{Gong_etal_2019} we model the laser pulse as a plane wave with potential ${\bm{a}}_\text{las}(t,\bm{x}) = \left|ma_0/e\right| \cos(\phi+\phi_0) \bm{e}_\text{las}$ with the dimensionless amplitude $a_0 = \left|eE_L\right|/m \omega$, $\phi=\omega(t-x_\|)$, and the initial phase $\phi_0$. In accordance with most experimental setups, we consider a laser with a pulse duration much longer than its optical cycle, whence we can approximate it to be monochromatic. Then, the energy $\varepsilon$ of an electron propagating at an angle $\theta \ll 1$ with respect to the channel axis, approximately changes as \cite{Gong_etal_2019}
\begin{align}
\frac{d\varepsilon}{dt} \approx m \omega a_0 \theta - \frac{\kappa \varepsilon^2 a_0^2  }4 \left[\theta^2 + \theta_B^2 \right]^2, \label{eq:Energy}
\end{align}
where $\theta_B := \sqrt{2a_\text{MB}/  a_0}$, $\kappa = 2 \omega^2 \alpha_\text{QED}/3m^2 \approx 4.5 \times 10^{-14}$ with the fine structure constant $\alpha_\text{QED}$, $a_\text{MB}=\left|e{\bm{a}}_\text{mag}(x_\text{MB})/m\right|$ is the maximum dimensionless channel potential experienced by the electron, and we dropped terms $\mathcal{O}\left(\gamma^{-2}\right)$. At a fixed energy the energy gain is maximal at an angle $\theta^\text{peak}$, which is the real root of the equation $\varepsilon = (m \omega /(\kappa a_0   \theta^\text{peak} [{\theta^\text{peak}}^2 + \theta_B^2])^{1/2}$. Using this expression in \cref{eq:Energy}, we find the maximal energy gain $d\varepsilon^\text{max}/dt \approx m \omega a_0 \left(3{\theta^\text{peak}}^2 - \theta_B^2 \right)/(4\theta^\text{peak})$, indicating that for $\theta \leq \theta^\text{min} := \theta_B/\sqrt{3}$ maximal energy gain is no longer possible. We will thus consider $\theta\gtrsim\theta_B$, in which case \cref{eq:Energy} is approximately maximal at
\begin{align}\label{eq:ThetaPeak}
\theta^\text{peak} \approx \left(\frac{m \omega}{\kappa \varepsilon^2 a_0 }\right)^{1/3}.
\end{align}
For propagation at this angle \cref{eq:Energy} transforms into a separable form with the solution
\begin{align}
    \varepsilon (t) &= \left(\frac{5 }{4} \left(\frac{m^4 \omega^4 a_0^2}{\kappa }\right)^{\frac13} \left(t-t_0\right) + \varepsilon_0^{\frac53} \right)^{\frac35}. \label{Eq:LargeAnglePeakEnergyGain}
\end{align}
On the other hand, the electron's radiative energy loss is approximately given by the second term in \cref{eq:Energy}. And since the energy $\varepsilon(\theta)$ is a monotonically decreasing function of the propagation angle, the emitted power has a nontrivial angular dependence which, for an electron with initial propagation angle $\theta_0$, features a minimum at 
\begin{align}
 \theta^* = \sqrt{\frac{\varepsilon_0\left(2-2\cos\left(\theta_0\right)+\theta_B^2\right)}{2\varepsilon} - \frac{\theta_B^2}{2}}. \label{eq:MinimalEmission}
\end{align}
This angle is always smaller than $\theta^\text{peak}$. As a consequence we see that particles will only gain significant energy if propagating at angles larger than $\theta^*$ and emit little energy into smaller angles. Hence, combining \cref{Eq:LargeAnglePeakEnergyGain,eq:MinimalEmission} we obtain $\theta^*(t)$ which constitutes a maximal angle with respect to the channel axis below which there will be only negligible emission. This prediction can now be compared to the angular distribution of the radiation obtained numerically. In doing so, however, it is important to note that the energy increase captured in \cref{Eq:LargeAnglePeakEnergyGain} drives a reduction in the emission angle $\theta^*$, i.e., a closing of the emission cone, only until the electron has reached a propagation angle $\theta^\text{min}$. Beyond that angle there can be no further energy gain and we estimate the continuing closing of the emission cone by estimating the time it takes until the electron's continuing emission overcomes the detection threshold $\varepsilon^\text{det}=100$ MeV, set for detection in the above PIC simulation. As we are interested in scaling laws, we approximate $\theta^\text{min} \approx \theta_B$ and estimate the peak energy an electron can attain maximally by inserting $\theta^\text{peak}=\theta^\text{min}$ into \cref{eq:ThetaPeak} to find $\varepsilon^\text{max} \approx  (m \omega /(\kappa^2 a_0 \theta_B^3))^{1/3}$. An electron cannot be accelerated to energies $\varepsilon\gtrsim\varepsilon^\text{max}$ in the plasma channel, as its radiative losses would overcompensate the laser's acceleration. In order to estimate the closing of the emission cone once the electron has reached this threshold we insert $\varepsilon^\text{max}$ into the emitted power and compute the difference $\Delta t$ of the times required to radiate an energy corresponding to the detection threshold into a given angle $\theta\leq\theta^\text{min}$ and into $\theta^\text{min}$. We then invert the resulting expression to find the angle at which the emission threshold is overcome after a given time $\Delta t$ has elapsed after the electron reached $\varepsilon^\text{max}$, given by
\begin{align}
\theta(\Delta t) = \theta_B \sqrt{\sqrt{\frac{4\varepsilon^\text{det}}{a_0 m \omega \theta_B \Delta t + \varepsilon^\text{det}}} - 1}. \label{eq:SmallAngle}
\end{align}

In contrast to these classical considerations, stochastic effects of quantum radiation will alter the emission characteristics. Replacing the expression for the emitted power in \cref{eq:Energy} by a sum over discrete photon emissions turns the electron's energy evolution into a stochastic differential equation, ultimately altering its dynamics from \cref{Eq:LargeAnglePeakEnergyGain}. Instead of repeating this full procedure, however, we note that a similar analysis has already shown that in the collision of an electron bunch with a laser pulse in vacuum, due to the emission's quantum stochasticity, the central light-cone momentum $p_- = \varepsilon - p_\|$ of an electron bunch's energy distribution is shifted with respect to its classical value $p_-^c$ according to \cite{Neitz_DiPiazza_2013,Neitz_Dissertation_2014}
\begin{align}
 p^q_-(t) = \frac{p_-^c(t)}{1 - p_{0,-} c_\alpha \Phi(t) - p_{0,-}^2 c_\beta \int dt' \frac{1}{1 - c_\alpha \Phi(t') p_{0,-}}},
 \label{eq:QEDSpreading}
\end{align}
with the initial light-cone momentum $p_{0,-} = \varepsilon_0(1-\cos(\theta_0))$, $\Phi(t) = \int dt (1-\cos[\theta^\text{peak}(t)])$ is the laser phase of the most energetic electrons and the numerical prefactors $c_\alpha=-\alpha_\text{QED}/(3\omega) (a_0 \omega/m)^2$, $c_\beta=\alpha_\text{QED}/(\sqrt{108}\pi m\omega) (a_0 \omega/m)^3$ are reduced with respect to literature values by the average of the second and third power of a harmonic oscillation over one oscillation period, to account for the monochromatically oscillating electric field, assumed here. 
%
\begin{figure}[t]
	\centering
	\includegraphics[width=\linewidth]{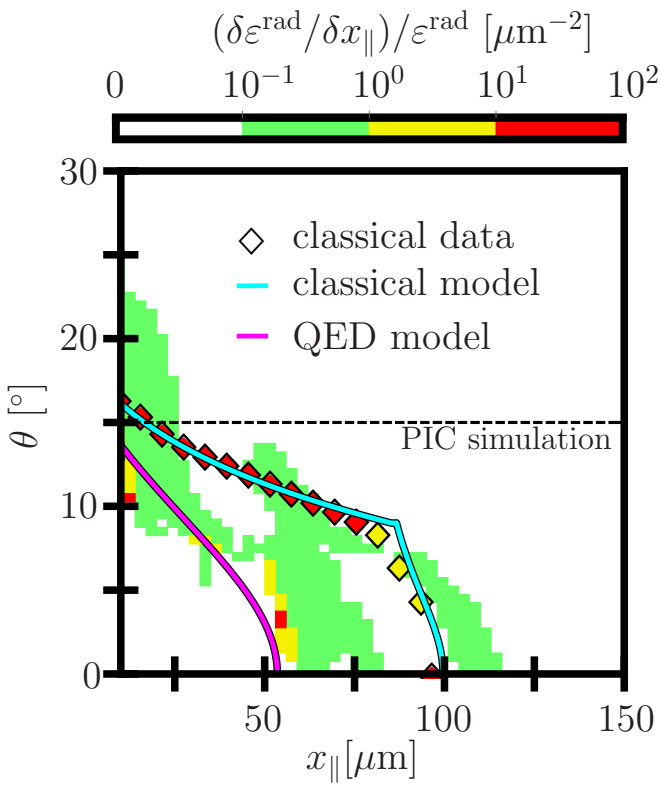}
	\caption{Closing of the emission cone with increasing channel length as observed in test particle simulations with continuous classical emission model (diamonds) and stochastic QED emission (background color). Model predictions \cref{eq:ThetaPeak} of minimal emission angle for an electron as a function of its longitudinal position inside the plasma channel for the classical (\cref{Eq:LargeAnglePeakEnergyGain}, cyan) and QED emission model (\cref{eq:QEDSpreading}, magenta).}
	\label{fig:QEDEnergyDependences}
\end{figure}
Since the transformation between light-cone momentum and energy is equivalent in a classical and quantum analysis, respectively, we approximate an electron's maximal energy inside the plasma channel according to the QED emission model as 
\begin{align}
\varepsilon^q(t) = \frac{p_-^q(t)}{1-\cos\left[\theta^\text{peak}(t)\right]}. \label{eq:QEDEnergyDynamics}    
\end{align}
Naturally, in this way, if we neglect stochastic effects in the emission, i.e., set $c_\alpha=c_\beta \equiv0$ in \cref{eq:QEDSpreading}, we recover the classical energy evolution $\varepsilon^c(t)$ according to \cref{Eq:LargeAnglePeakEnergyGain}. Since the denominator of \cref{eq:QEDSpreading} is always smaller than unity for the parameters we study, we will find $\varepsilon^q(t) \geq \varepsilon^c(t)$ at all times. Hence, inserting $\varepsilon^q(t)$ into the angle thresholds \cref{eq:MinimalEmission,eq:SmallAngle} we expect the emission cone to close at earlier times in the quantum emission analysis, as compared to the classical case. This prediction translates to a higher emission in the quantum case, which is in agreement with the conclusion drawn from the full PIC simulation (s. \cref{fig:figure2}).
We note, however, that in this PIC simulation the emission is given as a function of the electron's longitudinal position inside the plasma channel. In order to compare the model developed above, we derive this position as the integral over the electrons' effective longitudinal drift velocity. 
The exact longitudinal drift is difficult to compute, due to the highly nonlinear couplings in the problem. Hence, we approximate it as the drift velocity of an electron in a plane wave laser field with average propagation angle $\theta$ with respect to the laser's propagation direction, which is given by $v_\|^\text{eff}(\theta) = \cos\left(\theta\right)\left[1-\cos\left(\theta\right)\right]+\mathcal{Q}^2(t)/\left[1-\cos\left(\theta\right)+\mathcal{Q}^2(t)\right]$ with $\mathcal{Q}(t) = ma_0/2\varepsilon(t)$ \cite{Salamin_Faisal_1996}. The most energetic particles on average propagate along the angle of strongest energy gain $\theta^\text{peak}$. Thus, the longitudinal position can be approximated as $x_\|(t) = \rho_v \int_{0}^{t} dt' v^\text{eff}_\|(\theta^\text{peak}(t'))$, where we added the fitting parameter $\rho_v$ to account for effective reductions of the drift velocity due to the channel magnetic field. 

With this connection, we test the analytical prediction by running a full numerical simulation of the test particle model's equations of motion \cite{Gong_etal_2019} and compare these reduced numerics as well as the analytical model expressed in \cref{Eq:LargeAnglePeakEnergyGain,eq:MinimalEmission,eq:SmallAngle,eq:QEDEnergyDynamics} to the full PIC simulation. Furthermore, we have an experiment in mind where the discontinuous jump in the total emitted energy, observed in the classical analysis (s. red line \cref{fig:figure2}) and explained by our analysis as being due to the closing of the emission cone, is sought as the tell-tale signature of continuous classical emission. A potential experiment can be conducted by consecutively increasing the length of the laser-driven plasma channel target and measuring the total emitted energy. The laser is modelled with a dimensionless amplitude $a_0=270$, as used in the PIC simulation, and we model the channel magnetic field to have a dimensionless amplitude $\alpha=\pi^2/5$ closely resembling the channel current observed in the PIC simulation. We assume an electron bunch inside the plasma channel with initial energies $\varepsilon_0\in\left[80m,120m\right]$ and a typical longitudinal momentum $p_\| = 20m$ capturing the broad range of injection angles observed in the PIC simulation. Finally, since the initial stages of the acceleration are involved to model, and our model \cref{eq:ThetaPeak} does not hold for large propagation angles, we fix the initial conditions for our analytical model such that the initial angle $\theta_0$ is given by to the largest emission angle observed in the numerical integration of the test particle model equations and the initial energy $\varepsilon_0$ of our model such that $\theta^\text{peak}(\varepsilon_0)=\theta_0$, according to \cref{eq:ThetaPeak}. With these initial conditions we find that in the QED model a jump in the relative energy increase $\delta \varepsilon^\text{rad}/\delta x_\| \varepsilon^\text{rad}$ occurs at much lower channel lengths, i.e., the emission cone closes much faster than in the classical model (s. \cref{fig:QEDEnergyDependences}), which corresponds to higher particle energies in the channel and hence stronger emission.  Note that the color code of the classical and QED data is plotted on the same scale, indicating that the classical emission features a much sharper jump in the relative energy increase, as also observed in the PIC simulation. We can now attribute this behavior to the larger electron energies allowed when considering quenching of quantum emission \cite{Harvey_etal_2017}, effectively reducing the angle of emission, according to \cref{eq:ThetaPeak}. 
To fix the free effective propagation velocity of our analytical model, reduced due to the channel magnetic field, we ran a least-$\chi^2$ error scan, resulting in the optimal value $\rho_v\approx0.3$. We used this parameter to plot the solid lines in \cref{fig:QEDEnergyDependences}. We find excellent agreement of the numerical data with the model prediction of the minimal emission angle \cref{eq:MinimalEmission,eq:SmallAngle} in the classical as well as the quantum analysis, modeled by \cref{Eq:LargeAnglePeakEnergyGain,eq:QEDEnergyDynamics}, respectively. We note that the discontinuity of the classical model occurs at the point $\varepsilon(t) = \varepsilon^\text{max}$ and corresponds very well to a discontinuity in the classical data, obtained from the full numerical integration of the test particle model's equations of motion. Physically, this discontinuity is thus due to the fact that at this emission angle the mechanism responsible for closing the emission cone changes from an acceleration driven mechanism to an emission driven process, caused by the fact that electrons inside the plasma channel radiate more into larger angles. Naturally, this change of the emission mechanisms implies that after the discontinuity the relative jump in the total emitted energy is smaller than before the discontinuity, which is also confirmed by the jump in the classical relative energy increase being smaller  after the discontinuity (s. \cref{fig:QEDEnergyDependences}). In the quantum analysis, however, we note that due to $\varepsilon^q(t)\gg\varepsilon^c(t)$, the emission cone is fully closed at earlier times, whence the discontinuity is far less pronounced. We moreover note, that for the observation angle $\theta^*=15^\circ$, as modelled in the PIC simulation, our classical model predicts a channel length of $x_\| \approx 16\, \mu$m required to close the emission cone, in reasonable agreement with the full PIC simulation (s. \cref{fig:figure2}). For the quantum emission model, on the other hand, the emission cone for this observation angle is closed at much smaller channel lengths, corresponding to the absence of a jump in the total emitted energy observed in the PIC simulation. 

In summary, we have identified and characterized a novel approach to studying stochastic signatures of quantum radiation reaction in structured plasma channel targets. The next step will be the preparation of an experimental campaign for which purpose a systematic array of kinetic plasma simulations will have to be performed.


%

\end{document}